\documentclass[twocolumn,showpacs,preprintnumbers,amsmath,amssymb]{revtex4-1}
\usepackage{graphicx}
\usepackage{dcolumn} 
\usepackage{bm}      
\usepackage{xcolor}      
\usepackage[mathscr,mathcal]{euscript}

\begin{document}

\title{Reentrant quantum anomalous Hall effect with in-plane magnetic fields in HgMnTe quantum wells}
\author{Hsiu-Chuan Hsu$^1$, Xin Liu$^1$ and Chao-Xing Liu$^1$}
\affiliation{$^1$ Department of Physics, The Pennsylvania State University, University Park,
Pennsylvania 16802-6300}


\date{\today}

\begin{abstract}
Quantum anomalous Hall effect has been predicted in HgMnTe quantum wells with an out-of-plane magnetization of Mn atoms. However, since HgMnTe quantum wells are paramagnetic, an out-of-plane magnetic field is required to polarize magnetic moments of Mn atoms, which inevitably induces Landau levels and makes it difficult to identify the origin of the quantized Hall conductance experimentally.  
In this work, we study the quantum anomalous Hall effect in the presence of an in-plane magnetic field in Mn doped HgTe quantum wells. For a small out-of-plane magnetic field, the in-plane magnetic field can drive the system from a normal insulating state to a quantum anomalous Hall state. When the out-of-plane magnetic field is slightly above the transition point, the system shows a reentrant behavior of Hall conductance, varying from $-e^2/h$ to $0$ and back to $-e^2/h$, with increasing in-plane magnetic fields. The reentrant quantum anomalous Hall effect originates from the interplay between the exchange coupling of magnetic moments and the direct Zeeman coupling of magnetic fields. The calculation incorporating Landau levels shows that there is no qualitative change of the reentrant behavior.
\end{abstract}

 \maketitle
\section{Introduction}
When a two-dimensional (2D) electron gas moves in an external magnetic field perpendicular to the 2D plane, Lorentz force can induce a transverse current of electrons, known as the Hall effect\cite{hall1879a}. In 1980, K. von Klitzing discovered the quantum version of the Hall effect, the quantum Hall (QH) effect\cite{klitzing1980}, in which Hall conductance can be exactly quantized to an integer number in the unit $\frac{e^2}{h}$ due to the formation of Landau levels in strong magnetic fields. For a non-zero quantized Hall conductance, time reversal breaking is required, but strong magnetic fields, as well as Landau levels, are not necessary. In 1988, Haldane\cite{haldane1988} first proposed a theoretical model to realize the QH effect without Landau levels, which was mainly of academic interest\cite{Onoda2003} at that time and has been overlooked for almost twenty years. Recently, it is realized that the quantized Hall conductance can also be induced by the exchange coupling of magnetic moments\cite{Qi2006}, which provides a possibility to realize this effect in magnetic materials. Several realistic systems, including Mn doped HgTe quantum wells\cite{Liu2008b}, magnetic impurities doped Bi$_2$Se$_3$ thin films\cite{yu2010}, GdBiTe$_3$ thin films\cite{Zhang2011b}, {\it etc}\cite{Wu2008,Qiao2010, Zhang2012c}, have been proposed to possess the quantized Hall conductance. In analog to the anomalous Hall effect, where the Hall effect is induced by the exchange coupling of magnetic atoms in ferromagnetic conductors, the QH effect in these systems are dubbed as the ``quantum anomalous Hall'' (QAH) effect. After the successful discovery of topological insulators\cite{Hasan2010a,Qi2011}, a large experimental effort has been made to persue the realization of the QAH effect\cite{Chang2011, Zhang2012a, Buhmanna}, and recently the QAH effect has been realized in the Cr-doped (BiSb)$_2$Te$_3$ thin film\cite{Chang2013a}. 

The QAH effect was predicted in HgTe/CdTe quantum wells doped with magnetic ions Mn\cite{Liu2008b} when the magnetic moments are polarized along the out-of-plane direction. However, HgMnTe is a paramagnetic material, rather than a ferromagnetic material, a small magnetic field is required to polarize Mn magnetic moments. Consequently, the conventional QH effect due to Landau levels of magnetic fields coexists with the QAH effect induced by magnetization and it is difficult to identify the origin of the quantized Hall conductance because these two effects are topologically equivalent. Thus, it is desirable to find an experimentally feasible approach to distinguish these two effects. 

The orbital effect, as well as Landau levels, can only be induced by an out-of-plane magnetic field. In contrast, the direct Zeeman coupling of magnetic fields and the exchange coupling of magnetic moments exist for any direction of magnetic field. Therefore, the in-plane magnetic field provides a suitable tool to separate the exchange coupling or Zeeman coupling from the orbital effect of magnetic fields. Motivated by this idea, we study the influence of in-plane magnetic fields on the QAH effect in HgMnTe quantum wells. With increasing the in-plane magnetic field, a reentrant behavior appears in a certain regime of the out-of-plane magnetic field. The reentrant behavior of the QH states\cite{Zeitler2001a,Eisenstein2002a,Goerbig2003a} and other topological states\cite{Beugeling2012a} have been discussed in literatures and usually have quite different origins. In the present study, the reentrant QAH effect for the in-plane magnetic field occurs as a result of the competition between the exchange coupling of magnetic moments and the in-plane direct Zeeman coupling. This behavior is absent if there is no magnetic moments, so it can be viewed as a direct proof of the influence of magnetic moments on the QAH effect.

The paper is organized as follows: In Sec. II, the model Hamiltonian and the formalism are given. In Sec. III, we present the in-plane magnetization induced QAH effect at a fixed out-of-plane magnetization. In Sec. IV, the stability of the QAH effect in the presence of Landau levels (LL) is examined. Finally, we conclude with the discussion in Sec. VI. 

\section{model Hamiltonian}
In this section, we first introduce our model Hamiltonian for the Mn-doped HgTe quantum wells. The effective Hamiltonian
is written in the basis of $|E1+\rangle, |H1+\rangle, |E1-\rangle, |H1-\rangle$, with E1 and H1 denote electron and heavy hole sub-bands and
$\pm$ for opposite spin states. The form of the effective Hamiltonian is
given by \cite{Bernevig2006a}
\begin{eqnarray}
	&&H=H_{BHZ}+H_{m} \label{Ham-real}\\
&&H_{BHZ}=\epsilon({\bf k})+M({\bf k})\tau_z+A(k_x\sigma_z\tau_x-k_y\sigma_0\tau_y) \notag\\
&&\epsilon({\bf k})=C-Dk^2, \notag\\
&&M({\bf k})=m-Bk^2\notag 
\end{eqnarray}
,where the Pauli matrices $\tau$ denote the E1 and H1 states and $\sigma$ indicates the spin states. We denote the growth direction as the z-direction
and the quantum well plane as xy plane. The first term is the same as the effective model for HgTe quantum wells, first obtained by Bernevig, Hughes and Zhang, known as BHZ model \cite{Bernevig2006a}. 
The parameters $m$, $B$, $A$, $C$, $D$ in the BHZ model depend on the material details and can be found in Ref.\cite{Beugeling2012a,Buttner2011a}
$H_{m}$ describes the spin splitting of electron and hole sub-bands and its form is given by 
\begin{eqnarray}
	H_{m}= {\bf g}_{1}\cdot{\vec{\sigma}}\tau_0+{\bf g}_{2}\cdot{ \vec\sigma}\tau_z,\label{eq:Hmi}
\end{eqnarray}
, where ${\bf g}_{1}=\frac{1}{2}({\bf g}_{e}+{\bf g}_{h})$ and ${\bf g}_{2}=\frac{1}{2}({\bf g}_{e}-{\bf g}_{h})$. 
Here the vectors ${\bf g}_{e}$ (${\bf g}_{h}$) couples to electron (hole) spin and describe the spin splitting for the E1 (H1) sub-bands for
magnetic fields along different directions. 
There are two types of contribution for spin splitting, one from the direct Zeeman coupling of magnetic fields and the other 
from the exchange coupling to Mn doping, so the form of spin splitting is given by\cite{Liu2008b,Novik2005a} 
\begin{eqnarray}
	g_{e(h)i}=g_{e(h)i}^{zm}\mathcal{B}_i+g_{e(h)i}^{exc},\quad i=x,y,z. 
	\label{eq:spinsplitting}
\end{eqnarray}
The first term gives the Zeeman coupling with the g-factor $g_{e(h)i}^{zm}$ and the magnetic field $\mathcal{B}_i$ , while the second term describes the exchange coupling to Mn doping, given by 
\begin{eqnarray}
&&g_{e(h)i}^{exc}=\tilde{g}_{e(h)i}\langle S\rangle_i ,\quad i=x,y,z \\
&&\langle \bm{S} \rangle=-\bm{\hat{e}} S_0 B_{5/2} \left(\frac{5g_{Mn}\mu_B B }{2k_B(T+T_0)} \right) \label{Mnspin}
\end{eqnarray}
where $\tilde{g}_{e(h)i}$ is the coupling constant between electron (hole) band and Mn spin $S$. Eq.\ref{Mnspin} is the mean field approximation of the Mn magnetization and $\bm{\hat{e}}$ denotes the direction of the magnetic field,  $S_0=5/2$ is the Mn spin, $g_{Mn}=2$ is the g-factor of Mn , $T_0\approx 2.6K$ is to rescale the temperature to take into account the antiferromagnetic interaction between Mn ions\cite{Gui2004a} and $B_{5/2}$ is the Brillouin function.
In the spin splitting (\ref{eq:spinsplitting}), the Zeeman terms depends linearly on magnetic fields while the exchange coupling has a complicated non-linear dependence. Due to the quantum wells configuration, the g-factors $g^{zm}_{e(h)i}$ and $\tilde{g}^{exc}_{e(h)i}$ for the spin splitting are assumed to be isotropic in the xy plane, but different for the z direction. Without loss of generality, we only consider the x direction for the in-plane magnetic fields below. 
The parameters of the Hamiltonian (\ref{eq:Hmi}) can be found in Ref.\cite{Liu2008b, Beugeling2012a}. 
In the realistic systems, the in-plane spin splitting for the heavy hole sub-band depends on the cubic order of 
magnetic field $\mathcal{B}$ and magnetic moments $S$ \cite{Liu2013a}, which is neglected in the following ($\tilde{g}_{hx}=\tilde{g}_{hy}=g_{hx}^{zm}=g_{hy}^{zm}=0$). 

\section{The reentrant QAH effect}
The QAH effect in HgMnTe quantum wells with only z-direction magnetization has been investigated in the Ref.\cite{Liu2008b}. For zero in-plane magnetic field, it has been shown that the QAH phase can be realized in the regime $g_{ez}g_{hz}<0$ and $|g_{2z}|>|m|$, while it is a normal insulator when $|g_{2z}|<|m|$. In the following, we will investigate how the in-plane magnetic fields, as well as the in-plane magnetization, affect the phase diagram of this system. Since the quantized Hall conductance can only be changed when the bulk band gap is closed, two phases will share the same Hall conductance if they can be adiabatically connected without closing band gap. Therefore, one can identify the Hall conductance in a finite in-plane magnetic field by adiabatically connecting to the regime with zero in-plane magnetic field. 

\begin{figure}[htbp]
\begin{center}
\includegraphics[width=0.5\textwidth]{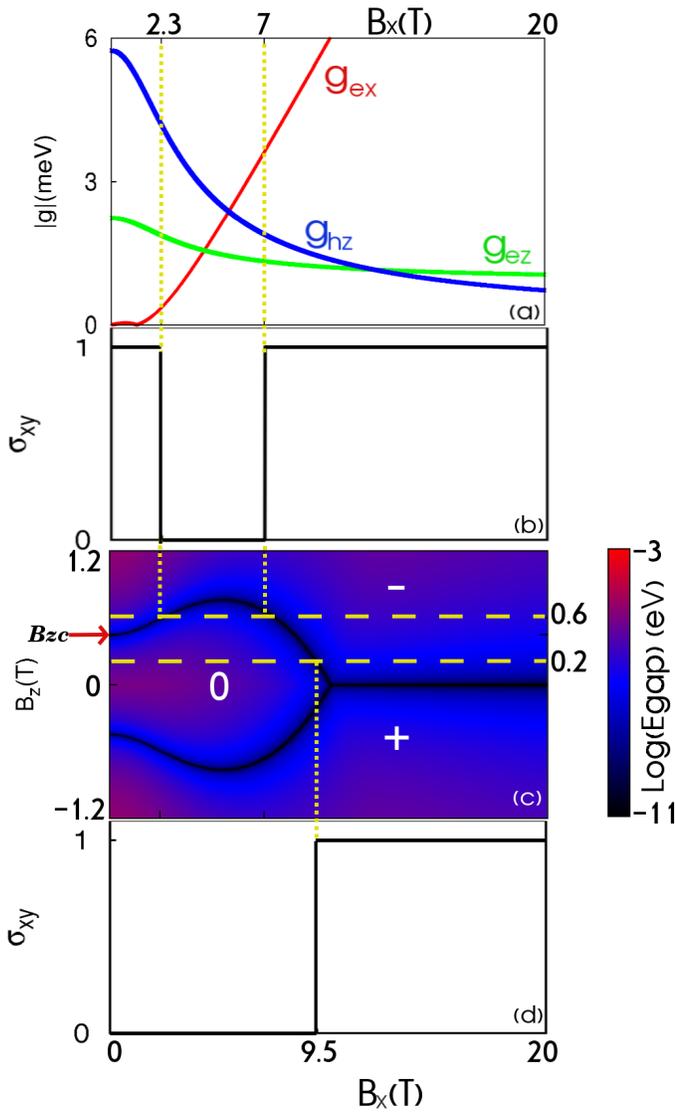}
\caption{(a)The effective spin splitting as a function of $B_x$ at a fixed $\mathcal{B}_z$=0.6T.  (b) The schematic plot of the Hall conductance in the unit of $e^2/h$ as a function of the $\mathcal{B}_x$ for the indicated $\mathcal{B}_z$=0.6T . (c)The phase diagram at 1K. The diagram is on a logarithmic scale to enhance the contrast. The black lines indicate the phase boundaries of different topologies.  (d)The schematic plot of the Hall conductance in the unit of $e^2/h$ as a function of the $\mathcal{B}_x$ for the indicated $\mathcal{B}_z$=0.2T.  The parameters used for Fig. (a) and (c) are $\tilde{g}_{ex}=-0.84 meV, g^{zm}_{ex}=-0.8meV/T, g_{hx}=0, \tilde{g}_{ez}=-2.13 meV, g^{zm}_{ez}=1.5meV/T,\tilde{g}_{hz}=9meV, g^{zm}_{hz}=-0.08meV/T, A=0.38 eV/nm, B=0.85 eV/nm^2, D=0.67eV/nm^2, m=3meV$.}
\label{main}
\end{center}
\end{figure}

We plot the energy gap for the Hamiltonian (\ref{Ham-real}) in Fig. \ref{main}(c) for $\mathcal{B}_x$ and $\mathcal{B}_z$ with the realistic parameters taken from Ref.\cite{Beugeling2012a} for HgMnTe quantum wells. The phase diagram is separated into three insulating regimes by the metallic lines, as depicted by the black lines in Fig. \ref{main}(c). For $\mathcal{B}_x=0$, the Hall conductance is known \cite{Liu2008b} to be $\pm\frac{e^2}{h}$ when $|\mathcal{B}_z|$ is larger than a critical value $|\mathcal{B}_{zc}|$ and zero when $|\mathcal{B}_z|<\mathcal{B}_{zc}$. Therefore, the Hall conductance of each insulating regime can be identified by adiabatic connection, as shown in Fig.\ref{main}(c). We find that with increasing $\mathcal{B}_x$, the critical z-direction magnetic field, as depicted by the metallic lines, first increases and then drops down to zero. For a fixed small $\mathcal{B}_z$ (the line $\mathcal{B}_z=0.2T$ in Fig.\ref{main}(c)), the system is driven from a normal insulator to a QAH insulator with Hall conductance $-e^2/h$ by increasing $\mathcal{B}_x$. More interestingly, when $\mathcal{B}_z=0.6T$, the Hall conductance $\sigma_{xy}$ undergoes the transitions from $-e^2/h$ to 0 to $-e^2/h$, showing a reentrant behavior for the QAH effect. This behavior is merely due to Zeeman coupling and exchange coupling rather than orbital effect since Landau levels are not considered in this calculation. In the following, we will discuss the physical picture of reentrant behavior of the QAH effect due to in-plane magnetic fields. 

\begin{figure}[htbp]
\begin{center}
\includegraphics[scale=0.5]{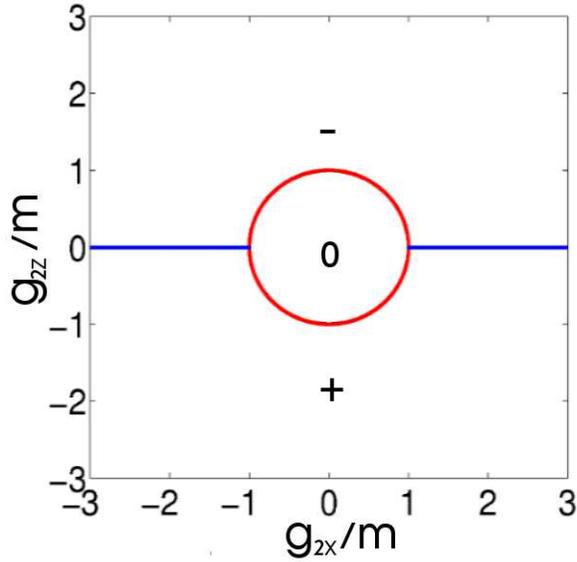}
\caption{The phase diagram obtained from analytical calculation where $g_{ex}=\pm g_{hx}$.}
\label{analyticmodel}
\end{center}
\end{figure}

To understand the transition, we first consider the case with either ${\bf g_{1}}=0$ or ${\bf g_{2}}=0$, in which the Hamiltonian can be solved analytically. In both cases, the results are qualitatively the same. Here, we show the result of ${\bf g_{1}}=0$. After diagonalization, the dispersion is
\begin{eqnarray}
E_{s,t}=s\sqrt{A^2k^2+M^2+g^2+t\sqrt{2A^2k_x^2g_{2x}^2+M^2g^2}} 
\end{eqnarray}
where $s,t=\pm$,  $k^2=k_x^2+k_y^2$, and $g^2=g_{2x}^2+g_{2z}^2$ is the strength of spin splitting. The energy gap is given by $E_{+-}-E_{--}=2\sqrt{A^2k^2+M^2+g^2 -\sqrt{2A^2k_x^2g_{2x}^2+M^2g^2}}$, which vanishes under the condition 
\begin{eqnarray}
[A^2k^2+M^2-g^2]^2+4A^2g_{2z}^2k^2+4A^2g_{2x}^2k_y^2=0. \label{cond}
\end{eqnarray}
This equation can be simplified as $g^2=m^2$  or $g_{2x}^2=A^2+(m-Bk^2)^2>m^2$ at $g_{2z}=0$. The gap-closing lines in terms of $g_{2x}$ and $g_{2z}$ are shown in Fig.\ref{analyticmodel}, separating three insulating phases. When $|g|>|m|$ and $g_{2z}\neq 0$, the system always stays in the QAH phase, regardless of the magnetization direction. The Hall conductances for positive and negative $\mathcal{B}_z$ have opposite signs\cite{Liu2008b}, which are separated by the metallic lines along $g_{2z}=0$ and $|g|>|m|$.

\begin{figure}[htbp]
\begin{center}
\includegraphics[width=0.5\textwidth]{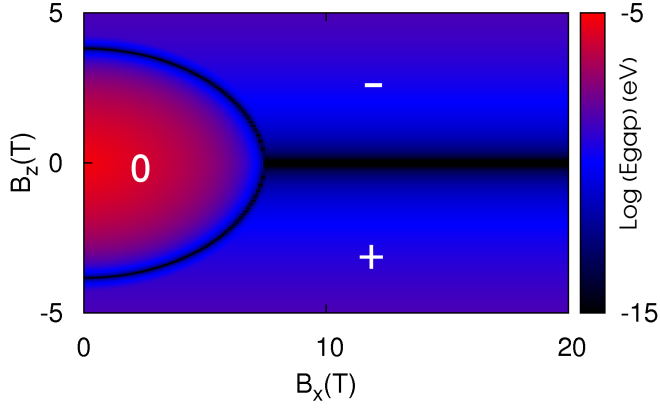}
\caption{The phase diagram at 1K for the Hamiltonian without the exchange coupling. The diagram is on a logarithmic scale to enhance the contrast. The black lines indicate the phase boundaries of different topologies.  The parameters used are $\tilde{g}_{ex}=0meV, g^{zm}_{ex}=-0.8meV/T, g_{hx}=0, \tilde{g}_{ez}=0meV, g^{zm}_{ez}=1.5meV/T,\tilde{g}_{hz}=0 meV, g^{zm}_{hz}=-0.08meV/T, A=0.38 eV/nm, B=0.85 eV/nm^2, D=0.67eV/nm^2, m=3meV$.}
\label{purezm}
\end{center}
\end{figure}

The analytic model suggests that the key factor for the normal insulator-QAH insulator transition is the strength of spin splitting $|g|$, instead of the direction of magnetic fields or magnetization. The magnetization direction does not have to be out-of-plane for the QAH effect to arise. For $\mathcal{B}_z=0.2T$, the spin splitting induced by z-direction magnetic field is not strong enough to induce the QAH state. With increasing the in-plane magnetic field, the total spin splitting is significantly enhanced, leading to the transition from the normal insulator to the QAH insulator at $\mathcal{B}_x=9.5T$, which is consistent with the above analytical solution. The reentrant behavior at $\mathcal{B}_z=0.6T$ results from the competition between the exchange coupling of magnetic moments and the direct Zeeman coupling of magnetic fields. For a small in-plane magnetic field $\mathcal{B}_x$, the exchange coupling is much stronger than the direct Zeeman coupling. So the spin splitting ${\bf g}_{e(h)}$ is dominated by the exchang term ${\bf g}^{exc}_{e(h)}$ and the direct Zeeman coupling part can be neglected. From the Kane model calculation, it turns out that the coupling constant of exchange coupling is strongly anisotropic\cite{Beugeling2012a,Buttner2011a}, and z-direction coupling is much stronger than the in-plane coupling. Consequently, when the magnetic moments of Mn atoms are tilted into x-direction due to the increase of $\mathcal{B}_{x}$, the spin splitting is reduced significantly, which leads to the transition from the QAH phase to the normal insulating phase. 
With further increasing in-plane magnetic field, the direct Zeeman term, which grows linearly with $\mathcal{B}_x$, is eventually dominant over the exchange term, which saturates at high magnetic fields. Thus, the system is driven back to the QAH phase. To verify this physical picture, we plot the spin splitting of E1 and H1 sub-bands as a function of $\mathcal{B}_x$ in Fig.\ref{main}(a). The green and blue curves show the rapid reduction of the z-direction spin splitting of E1 and H1 sub-bands respectively, as $\mathcal{B}_x$ increases. As a result, it leads to the transition from a QAH state to a normal insulating state at $\mathcal{B}_x\approx2.3T$. The red curve shows the growth of x-direction spin splitting and it eventually leads to the transition from normal insulating to QAH phase at $\mathcal{B}_x=7 T$. We would like to emphasize that the reentrant behavior is unique for the HgTe quantum wells with Mn doping. Without Mn doping, there is no exchange coupling to magnetic moments and consequently, we only find the transition from the normal insulator to the QH insulator regime, which is shown in Fig \ref{purezm}. 

For all the calculations above, the g-factors were estimated from Kane model calculation, which, to the best of our knowledge, have never been carefully identified in experiments. The anisotropy of the hole g-factor has been shown experimentally in p-type bulk HgMnTe\cite{Zverev1984}, which is consistent with parameters estimated from the Kane model. The qualitative picture of the reentrant behavior is independent of the parameter details.

\section{Landau levels with the in-plane magnetization}
\begin{figure}[htbp]
\begin{center}
\includegraphics[width=0.5\textwidth]{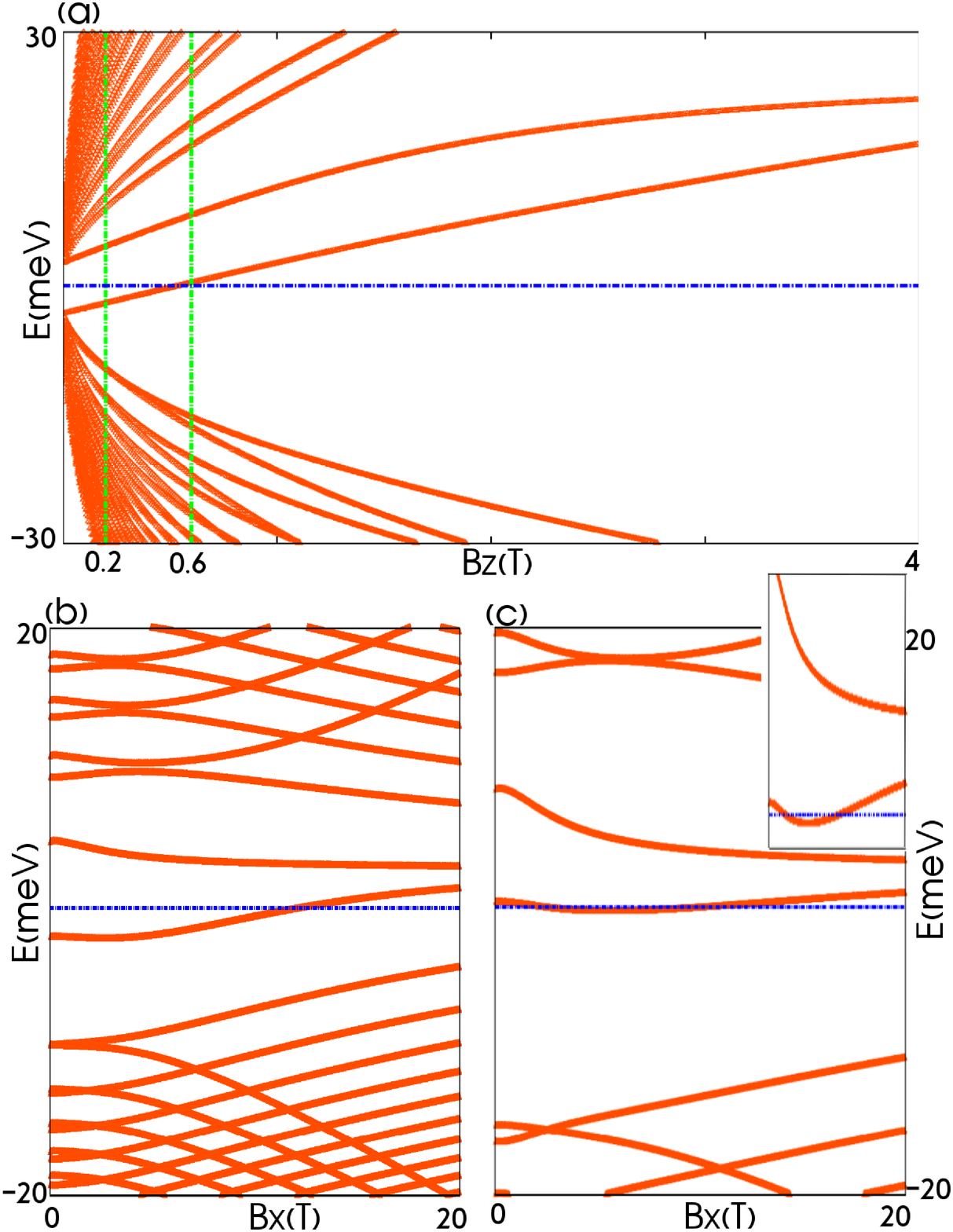}
\caption{Landau Level fan chart. The parameters are the same as in Fig.\ref{main}. The blue line indicate the Fermi level at $0.3$meV. (a)Landau level fan chart without in-plane magnetization. (b)Landau level  at $B_z=0.2T$ in terms of in-plane magnetic field. ($c$)Landau level  at $B_z=0.6T$ in terms of in-plane magnetic field. The inset zooms in near Fermi level and shows the reentrant behavior. }
\label{llx}
\end{center}
\end{figure}

In the above, we show the phase diagram of the QAH effect with both the in-plane and out-of-plane magnetic fields and find a novel reentrant behavior due to the combination of the exchange coupling and the direct Zeeman coupling. However, due to the non-zero $\mathcal{B}_z$, the formation of Landau level is inevitable. Therefore, it is natural to examine whether the reentrant behavior still exists after taking into account the orbital effect of Landau levels. Landau levels can be calculated by taking into account the orbital effect of magnetic fields in the model Hamiltonian (\ref{Ham-real}) with the standard Peierls substitution \cite{Novik2005a,Peierles1933}, which is described in details in the appendix. The Landau level fan chart is plotted in Fig.~\ref{llx}, with the Fermi level set at 0.3 meV (the blue line). Fig.~\ref{llx} (a) shows the Landau levels without $\mathcal{B}_x$, while Fig.~\ref{llx} (b) and ($c$) show how the Landau level evolves with $\mathcal{B}_x$ at $\mathcal{B}_z=0.2T$ and $\mathcal{B}_z=0.6T$, respectively. In Fig.~\ref{llx} (b), the system stays in the normal insulating regime for zero $\mathcal{B}_x$, and is driven to the regime with $-e^2/h$ with increasing $\mathcal{B}_x$, similar to the line $\mathcal{B}_z=0.2T$ in Fig.\ref{main}(c).  For $\mathcal{B}_z=0.6$T, before turning on $\mathcal{B}_x$, the Hall conductance is $\sigma_{xy}=\frac{-e^2}{h}$. The Fermi level crosses the electron zero mode twice and the Hall conductance undergoes the transitions from $-e^2/h$ to 0 to $-e^2/h$ as increasing $\mathcal{B}_x$. Fig.~\ref{llx} demonstrates the stability of the phase diagram given in Fig.~\ref{main}  in the presence of Landau levels and the underlying reason for the reentrant behavior is the change of spin splitting, rather than the orbital effect.

\section{conclusion}
In conclusion, we have shown that the in-plane magnetic field induces the QAH effect in HgMnTe quantum wells. A reentrant QAH effect is predicted as a result of the interplay between the exchange coupling and the direct Zeeman coupling. In addition, the reentrant behavior is stable in the presence of Landau levels, so it is feasible under the present experimental condition to verify this effect in HgTe quantum wells doped with Mn. 

\section{acknowledgments}
We would like to thank X.L. Qi for useful discussions. 

\section{appendix}

In this appendix, we show the Landau level calculation in the presence of in-plane magnetization. The calculation follows the perturbation theory in the reference, for example,~\cite{Winkler2003a}.  The full Hamiltonian is $H_{BHZ}+H_{m}$, where $H_{m}=H_{mx}+H_{mz}$ and $H_{mx(z)}={\bf g}_{1x(z)}{\vec\sigma_{x(z)}}\tau_0+{\bf g}_{2{x(z)}}{ \vec\sigma_{x(z)}}\tau_z$.
$H_{mx}$ is the in-plane magnetization and is regarded as the perturbation to Landau levels.
The unperturbed Hamiltonian $H_{BHZ}+H_{mz}$ with the standard Peierls substitution \cite{Novik2005a,Peierles1933} is
\begin{eqnarray}
H_o=H_{BHZ}+H_{mz}=\left(
\begin{array}{cc}
H_{o\uparrow} & 0 \\
0 & H_{o\downarrow}
\end{array}
\right) 
\end{eqnarray}
, where the Hamiltonian for each spin-component is a 2 by 2 matrix.
\begin{widetext}
\begin{eqnarray}
H_{o\uparrow}&=\left(\notag 
\begin{array}{cc}
C+M-\frac{2(B+D)}{l^2_B}(a_+a_-+\frac{1}{2})+g_{ez} &{\frac{\sqrt{2}A}{l_B}}a_+  \\
\frac{\sqrt{2}A}{l_B}a_-& C+M-\frac{(B-D)}{l_B^2}(a_+a_-+\frac{1}{2})+g_{hz} \\
\end{array}
\right)\\
H_{o\downarrow}&=\left(\notag 
\begin{array}{cc}
C+M-\frac{(B+D)}{l_B^2}(a_+a_-+\frac{1}{2})-g_{ez} &  -\frac{\sqrt{2}A}{l_B}a_ -\\
 -\frac{\sqrt{2}A}{l_B^2}a_+ &  C+M-\frac{(B-D)}{l_B^2}(a_+a_-+\frac{1}{2})- g_{hz}
\end{array}
\right)
\end{eqnarray}
\end{widetext}
, where $l_B=\sqrt{\frac{\hbar}{e\mathcal{B}_z}}$ is the magnetic length. 

$H_o$ is block-diagonal, while  $H_{mx}$ is off block diagonal. First, we calculated the eigenenergy and eigenstates for $H_o$. The eigenstates are written as
\begin{equation}
|n,l\rangle=\left(
\begin{array}{c}
f_{nl1}|n\rangle\\
f_{n-1l2}|n-1\rangle\\
f_{n-1l3}|n-1\rangle\\
f_{nl4}|n\rangle
\end{array}
\right)
\end{equation}
where $f_{nlj}$ ($j=1,2,3,4 $) are the coefficients of each eigenlevel of the Harmonic oscillator. $n$ denotes the eigenlevel of the Harmonic oscillator, $l=1,2,3,4,$ denotes the eigenstates of the effective $4\times 4$ Hamiltonian, and j denotes the components of eigenvectors.

For convenience, we define
\begin{eqnarray}
g_s=\frac{g_{hz} + g_{ez}}{2} \\
g_a=\frac{g_{hz} - g_{ez}}{2}
\end{eqnarray}
In this basis, the eigenvalues for the zero modes are 
\begin{eqnarray}
E_{\uparrow,0}&=&C+M-\frac{(B+D)}{l_B^2} +g_{ez}\\
E_{\downarrow,0}&=&C-M+\frac{(B-D)}{l_B^2} -g_{hz}\\
\end{eqnarray}

The eigenvalues for the non-zero modes are
\begin{widetext}
\begin{eqnarray}
E_{\uparrow,n\pm}&=&C-\frac{(B+2nD)}{l_B^2}+g_s 
\pm \sqrt{(M-g_a-\frac{(2nB+D)}{l_B^2})^2+A^2\frac{\sqrt{2n}}{l_B}}\\
E_{\downarrow,n\pm}&=&C+\frac{(B-2nD)}{l_B^2}-g_s
\pm \sqrt{(M+g_a-\frac{(2nB-D)}{l_B^2})^2+A^2\frac{\sqrt{2n}}{l_B}}
\end{eqnarray}
\end{widetext}
The eigenvectors for zero-modes are

\begin{eqnarray}
|\uparrow,0\rangle= \left(
\begin{array}{c}
|0\rangle\\
0\\
0\\
0
\end{array}
\right);
|\downarrow,0\rangle= \left(
\begin{array}{c}
0\\
0\\
0\\
|0\rangle
\end{array}
\right)
\end{eqnarray}
\newpage

The eigenvectors for non-zero modes are
\begin{widetext}
\begin{eqnarray}
|\uparrow,n\pm\rangle= \frac{-l_B}{\sqrt{2n} A}\left(
\begin{array}{c}
-M+g_a+\frac{e}{\hbar}(D+2nB)\pm
\sqrt
{
(M-g_a-\frac{e(2nB+D)}{\hbar})^2+2A^2\frac{\sqrt{2n}}{l_B})
}|n\rangle\\
|n-1\rangle\\
0\\
0
\end{array}
\right) \\
|\downarrow,n\pm\rangle= \frac{l_B}{\sqrt{2n}A}\left(
\begin{array}{c}
0\\
0\\
-M-g_a-\frac{e}{\hbar}(D-2nB)\pm
\sqrt
{
(M+g_a+\frac{e(D-2nB)}{\hbar})^2+2A^2\frac{\sqrt{2n}}{l_B})
} |n-1\rangle\\
|n\rangle
\end{array}
\right)
\end{eqnarray}
\end{widetext}
The positive sign is for the hole eigenstates, while the negative sign is for the electron states.
In this basis, $H_0$ is diagonal. Then we project $H_{mx}$ onto this basis and the total Hamiltonian is diagonalized up to 20 Landau levels numerically.

%

\end{document}